\documentclass[a4paper,11pt]{article}
\pdfoutput=1 

\usepackage{jheppub} 
\usepackage{slashed}
\usepackage{amsfonts}
\usepackage{amsmath}
\usepackage{amssymb}
\usepackage{bbm}
\usepackage[vcentermath]{youngtab}
\usepackage[T1]{fontenc} 
\usepackage{silence}

\newcommand{\beq}{\begin{eqnarray}}
\newcommand{\eeq}{\end{eqnarray}}

\newcommand{\bmp}{\noindent\begin{minipage}{16cm}}
\newcommand{\emp}{\end{minipage}\vskip 7mm} 

\newcommand{\SU}{\mbox{SU}}

\newcommand{\UU}{\mbox{U}}

\definecolor{pumpkin}{rgb}{1.0, 0.4, 0.0}

\title{Bridging the $ \mu $Hz Gap in the Gravitational-Wave Landscape: Unveiling Dark Baryons}

\author[a]{Martin Rosenlyst}

\affiliation[a]{Rudolf Peierls Centre for Theoretical Physics, University of Oxford, Parks Road, Oxford OX1 3PU, United Kingdom\\}

\emailAdd{martin.jorgensen@physics.ox.ac.uk}

\abstract{We study gravitational waves (GWs) with frequencies in the $\mu$Hz range, which arise from phase transitions related to dark confinement in the context of dark versions of Quantum Chromodynamics. Based on several compelling motivations, we posit that these theories predict the existence of GeV-mass asymmetric dark baryons, akin to ordinary baryons, with the potential to contribute to dark matter.
Furthermore, we emphasize the significance of a particular $\mathcal{O}(\text{TeV})$ scale for multiple reasons. First, to account for the similarity in present-day mass densities between dark matter and visible matter, various TeV-scale mechanisms can elucidate the similarities in both their number densities and masses. Moreover, to address the so-called electroweak hierarchy problem, we consider the introduction of either the Composite Higgs or Supersymmetry at around $\mathcal{O}(\text{TeV})$. These mechanisms lead to intriguing TeV collider signatures and the possibility of detecting mHz GWs in future experiments. In summary, this study provides a strong motivation for advancing GW experiments that can bridge the $\mu$Hz frequency gap in the GW spectrum. Additionally, there is a need for the construction of more powerful particle colliders to explore higher energy regimes. In consideration of the possibility to scrutinize these models from various perspectives, we strongly advocate their further development.}

\begin{document} 
\maketitle
\flushbottom

\section{Introduction}

Astronomical and cosmological observations~\cite{Planck:2018vyg} provide compelling evidence pointing towards Dark Matter (DM) as the dominant form of matter in the Universe, constituting around 85\% of its total mass. Understanding the non-baryonic nature of DM remains one of the most profound enigmas in our comprehension of the physical world. 
In the framework of the standard Cold DM (CDM) cosmology, the analysis of data from the Planck space telescope has yielded a precise determination of the mass density of CDM: $ \Omega_{\rm DM} h^2 = 0.1198\pm 0.0012 $~\cite{Planck:2018vyg}. This value corresponds to a ratio of DM to baryon mass density, $ \Omega_{\rm DM}/\Omega_{\rm B} $, equal to $ 5.36 \pm 0.10 $, where $ \Omega_X $ represents the mass density $ \rho_X $ of a species denoted as $ X $ divided by the critical density $ \rho_c $. Weakly interacting massive particles (WIMPs) have emerged as the leading candidates to explain the observed energy density of CDM~\cite{Feng:2010gw}. The masses of these particles are connected to the electroweak (EW) scale, and their abundance or relic density arises from a thermal freeze-out mechanism.

In this paper, we interpret this similarity ($ \Omega_{\rm DM}\simeq 5 \Omega_{\rm VM} $) between the present-day mass densities of DM and visible matter (VM), referred to as the cosmological coincidence, to be a hint towards a fundamental connection between the origins of DM and VM abundances. However, majority of prominent DM candidates do not provide a clear explanation for this relationship. The number density of VM is attributed to the baryon asymmetry of the universe, while its mass arises from the confinement energy of Quantum Chromodynamics (QCD). Given these distinct mechanisms for generation, there is no inherent reason why the cosmological mass densities of these species should align within the same order of magnitude. This coincidence in the mass densities largely motivates a DM particle-antiparticle asymmetry, similar to what is observed for VM. These related asymmetries likely arose from specific processes during an early cosmological epoch that subsequently decoupled, resulting in similar number densities for VM and DM --- $ n_{\rm VM} \sim n_{\rm DM} $. This is known as the Asymmetric DM (ADM) hypothesis. 

To adequately explain the cosmological coincidence, it is crucial to provide a rationale for the similarity in particle masses between VM and DM, $ m_{\rm VM} \sim m_{\rm DM} $. This is essential as the mass density can be expressed as $\Omega_{X} = n_{X} m_{X}/\rho_c$, and understanding the similarity in masses between VM and DM becomes necessary. By establishing a connection between the two, we can fully address the cosmological coincidence problem. Surprisingly, this aspect is often overlooked in the majority of literature focusing on ADM. This observation provides motivation for considering the possibility that DM particles are composite states, consisting of more fundamental constituents. This notion draws parallels to protons and neutrons, which consist of confined quarks. This suggests the intriguing possibility that DM particles may be formed through a dark analog of QCD at a confinement scale akin to the one observed in QCD. Moreover, the presence of an $ \UU(1) $ baryon symmetry in QCD ensures the stability of the lightest baryon, the proton, against decays. This compellingly supports the idea that DM might exist as dark baryons, stabilized by a similar symmetry, enabling them to achieve a relic density. Finally, in composite DM models incorporating strong confining gauge forces, we anticipate a natural propensity for composite DM to exhibit strong self-interactions capable of alleviating the small-scale structure problems of the collisionless CDM paradigm, as discussed in Refs.~\cite{Cline:2013zca,Cline:2022leq}.

This paper presents an overview of three model frameworks aimed at addressing the relationship between VM and DM masses, offering an explanation for the cosmological coincidence. Inspired by the proton analogy in QCD, we investigate ADM candidates known as dark baryons, which manifest as confined states of a dark QCD-like gauge group. Previous attempts to tackle this issue have followed different approaches. 
Firstly, we consider models involving the introduction of a mirror symmetry between the visible and dark gauge groups, commonly referred to as mirror matter models. In these models, this symmetry is either preserved~\cite{Hodges:1993yb,Foot:2003jt,Foot:2004pq,Foot:2004pa,Ibe:2018juk} or deliberately broken~\cite{An:2009vq,Cui:2011wk,Akamatsu:2014qsa,Farina:2016ndq,Lonsdale:2017mzg,Lonsdale:2018xwd,Ibe:2019ena,Beauchesne:2020mih,Feng:2020urb,Ritter:2021hgu,Ibe:2021gil,Bodas:2024idn}, such as Twin Higgs models~\cite{Farina:2015uea,Craig:2015xla,GarciaGarcia:2015pnn,Barbieri:2016zxn} that address the so-called little hierarchy problem. Secondly, we entertain the possibility that an extension of the field content can result in the convergence of the gauge couplings towards infrared fixed points. This establishes a connection between the confinement scales of visible and dark QCD, which yields the relationship $ m_{\rm VM} \sim m_{\rm DM} $, as initially proposed in Ref.~\cite{Bai:2013xga}. Lastly, we review the concept of dark unification, as proposed in Ref.~\cite{Murgui:2021eqf}, which presents a framework that unifies the dark and visible sectors. This establishes a link between the masses of dark and visible baryons through matching confinement scales. Interestingly, these model frameworks typically involve particles at the TeV scale, with masses ranging from $\mathcal{O}(1)$ to $\mathcal{O}(10)$~TeV. As discuss in these works, the detection of these particles may become possible in future collider experiments.

Furthermore, we motivate to elucidate the similarity between visible and dark baryon asymmetries in the early Universe, specifically $ n_{\rm VM}\sim n_{\rm DM} $. Low-scale leptogenesis realizations involving neutrino mass generation, such as resonant leptogenesis~\cite{Pilaftsis:1997jf,Pilaftsis:2003gt,Suematsu:2011va} and thermal leptogenesis in radiative neutrino mass models~\cite{Ma:2006km,Cai:2017jrq,Cai:2017jrq}, offer promising explanations for this coincidence. These models are further motivated by the existence of new physics at around the TeV scale. Finally, the need to address the EW hierarchy problem and explain the observed Higgs mass without fine-tuning has spurred exploration of various theories. These include Natural Supersymmetry (SUSY)~\cite{Batra:2003nj,Ellwanger:2009dp,Hall:2011aa,Curtin:2014zua,Casas:2014eca}, Composite Higgs (CH)~\cite{Kaplan:1983sm}, and proposals such as the Little Higgs~\cite{Arkani-Hamed:2001nha,Schmaltz:2005ky}. However, in the case of Little Higgs, it stabilizes the Higgs sector up to an experimental cutoff of $ 5-10 $ TeV, with additional physics ensuring naturalness at higher scales. These frameworks anticipate the presence of TeV-scale particles, particularly colored partner particles, that have not yet been observed at the Large Hadron Collider (LHC), underscoring the necessity for more powerful colliders.

The three model frameworks mentioned above, developed to address the mass-density coincidence problem, propose DM candidates as dark baryons with masses on the order of GeV. Given the outlined reasons, this particular scale of DM could potentially be associated with the presence of new physics within the range of $ 1-10 $~TeV. Exploring this correlation offers promising avenues to address various phenomena, including addressing the mass-density coincidence problem, providing a natural explanation for the Higgs mass, generating asymmetries and elucidating the origin of the neutrino masses. Therefore, we strongly encourage the development of models that comprehensively explore these possibilities or other potential explanations. 

The primary objective of this paper extends beyond motivating the exploration of new physics within the range $1-10$ TeV using future colliders. It specifically focuses on the potential generation of gravitational wave (GW) signals in dark baryon models. Our study highlights that if the phase transition triggered by dark confinement, occurring around $ \mathcal{O}(1) $~GeV, is of first order, it has the potential to generate GWs spanning a broad frequency range, from $\mu$Hz to mHz. Remarkably, the field of GW research has, to date, left the frequency band of $\mu$Hz largely unexplored. This crucial consideration provides compelling grounds to pursue the development of GW experiments capable of bridging the frequency gap in the GW landscape, with a particular emphasis on the $ \mu $Hz range. Consequently, it serves as a strong motivation for advancing research and technology in the field of GWs.

\section{Dark baryons}
\label{sec: Dark baryons}

We assume the existence of a dark sector analogous to QCD, consisting of $N_{q_d}$ flavors of light dark quarks. These dark quarks have masses well below the confinement scale $ \Lambda_{\rm dQCD} $, enabling us to treat the dark pions as pseudo-Nambu-Goldstone bosons (pNGBs). Within this dark sector, composite DM candidates emerge, resembling nucleons in the visible sector. These candidates are bound states of dark quarks, known as dark baryons, confined by a new $\SU(N_d)$ gauge symmetry. We suggest that the masses of these dark baryons are similar to those of protons and neutrons, indicating a correspondence between the confinement scales of the dark sector and QCD. For simplicity, we will assume that all quarks have the same mass, denoted as $ m_{q_d} $, and that there are three colors and three light flavors ($ N_{q_d}=N_d=3 $), mirroring the framework of QCD. A naive estimation, in order to align with the mass of visible nucleons, would suggest that the mass of the dark nucleons are approximately $m_{N_d}\simeq 3.8 \Lambda_{\rm dQCD}$, while the dark pions have the mass $ m_{\pi_d}\propto \sqrt{m_{q_d}\Lambda_{\rm dQCD}} $. 

\subsection{Dark matter relic density}
\label{sec: Dark matter relic density}

Similar to VM, dark nucleons also have a conserved number, indicating that they may be ADM. The origin of this asymmetry, which is also unknown for visible baryons, likely arises from physics at higher scales than that of DM, potentially at $ \mathcal{O}(1-10) $~TeV as discussed in Section~\ref{sec: Generation of asymmetries}. 

However, a question arises regarding the relic density of dark pions, which are necessarily present. One possibility is that the dark quarks are massless, resulting in the pions being true Goldstone bosons and contributing solely to the dark radiation density in the universe. Alternatively, if the pions have mass, the existence of massless dark photons that couple to the hidden quarks could facilitate efficient annihilation. However, this introduces a new long-range interaction between dark nucleons, potentially complicating the viability of DM candidates and necessitating additional species to ensure the $ \UU(1)^\prime $ charge neutrality of the universe. A less problematic approach involves assigning masses greater than twice the electron mass ($ 2m_e $) to the dark photons ($ \gamma^\prime $), ensuring that $ \pi_d \pi_d \rightarrow \gamma^\prime \gamma^\prime $ annihilation remains efficient. Simultaneously, through kinetic mixing of $ \gamma^\prime $ with photons, the $ \gamma^\prime $ particles can decay into electron-positron pairs ($ \gamma^\prime \rightarrow e^+e^- $). For a more comprehensive discussion of the constraints related to relic density, readers are directed to Ref.~\cite{Cline:2022leq}. 

\subsection{Dark matter self-interactions}
\label{sec: Self-interactions}

Given our assumption that the masses of dark quarks are significantly lower than $\Lambda_{\rm dQCD}$, we can treat dark pions as pNGBs, enabling us to utilize the determinations of nucleon scattering lengths $a$ obtained from lattice data in Ref.~\cite{Chen:2010yt}. The S-wave nucleon-nucleon scattering amplitude is given by~\cite{Chu:2019awd} \begin{equation} \begin{aligned} \label{eq: N-N scattering amplitude}
\mathcal{A}=\frac{4\pi}{m_N(-ip-a^{-1}+r_0p^2/2+\mathcal{O}(p^4))}\,,
\end{aligned}\end{equation} where $ m_N $ is the nucleon mass, $ p $ is the center-of-mass momentum, $ r_0 $ is the effective range, while the scattering length $ a $ can have an effect on the velocity-dependence of the cross section at low energy. The self-interacting cross section can be written as \begin{equation} \begin{aligned}
\sigma = \frac{m_N^2}{16\pi}\left(\vert \mathcal{A}_s\vert^2+3\vert \mathcal{A}_t\vert^2\right)\,,
\end{aligned}\end{equation} where $ \mathcal{A}_s $ and $  \mathcal{A}_t  $ represent the scattering amplitudes given in Eq.~(\ref{eq: N-N scattering amplitude}) in the singlet and triplet spin channels, respectively. In Ref.~\cite{Chen:2010yt} by using the nuclear effective field theory~\cite{Beane:2008bt}, the scattering lengths are derived in the singlet and triplet spin channels for $ \Lambda_{\rm QCD}=250 $~MeV as follows \begin{equation} \begin{aligned}
a_s =\frac{0.71 \Lambda_{\rm QCD}^{-1}}{m_\pi/\Lambda_{\rm QCD}-0.58}\,, \quad\quad a_t =\frac{0.96 \Lambda_{\rm QCD}^{-1}}{m_\pi/\Lambda_{\rm QCD}-0.42}\,.
\end{aligned}\end{equation} Based on the findings presented in Figure 2 of Ref.~\cite{Chen:2010yt}, the values of $ r_0 $ are considerably smaller compared to the scattering lengths $ a $ within the relevant parameter space. This allows us to neglect $ r_0 $ without significantly affecting the scattering results.

The DM self-scattering cross section $\langle \sigma v\rangle$ is computed in Ref.~\cite{Cline:2022leq} by performing a phase space average over the DM velocity distribution. By assuming a Maxwellian distribution, expressing momentum in terms of the relative velocity as $p=m_N v/2$ and imposing the condition $r_0\ll a$, it follows that \begin{equation} \begin{aligned} \label{eq: velocity-weighted CS}
 \langle \sigma v\rangle = \frac{a_s F(b_s)+3a_t F(b_t)}{m_N}\,, \quad \quad
 F(b)=\sqrt{\frac{\pi}{2}}\left(2b+b^3 e^{b^2/2}\text{Ei}(-b^2/2)\right).
\end{aligned}\end{equation}Here, Ei is the exponential integral function and $ b_i\equiv 4/(\sqrt{\pi}m_N a_i \langle v\rangle) $ with $ i=s,t $, while $ \langle v\rangle $ is the mean DM velocity. 

\begin{figure}[t]
	\centering
	\includegraphics[width=0.8\textwidth]{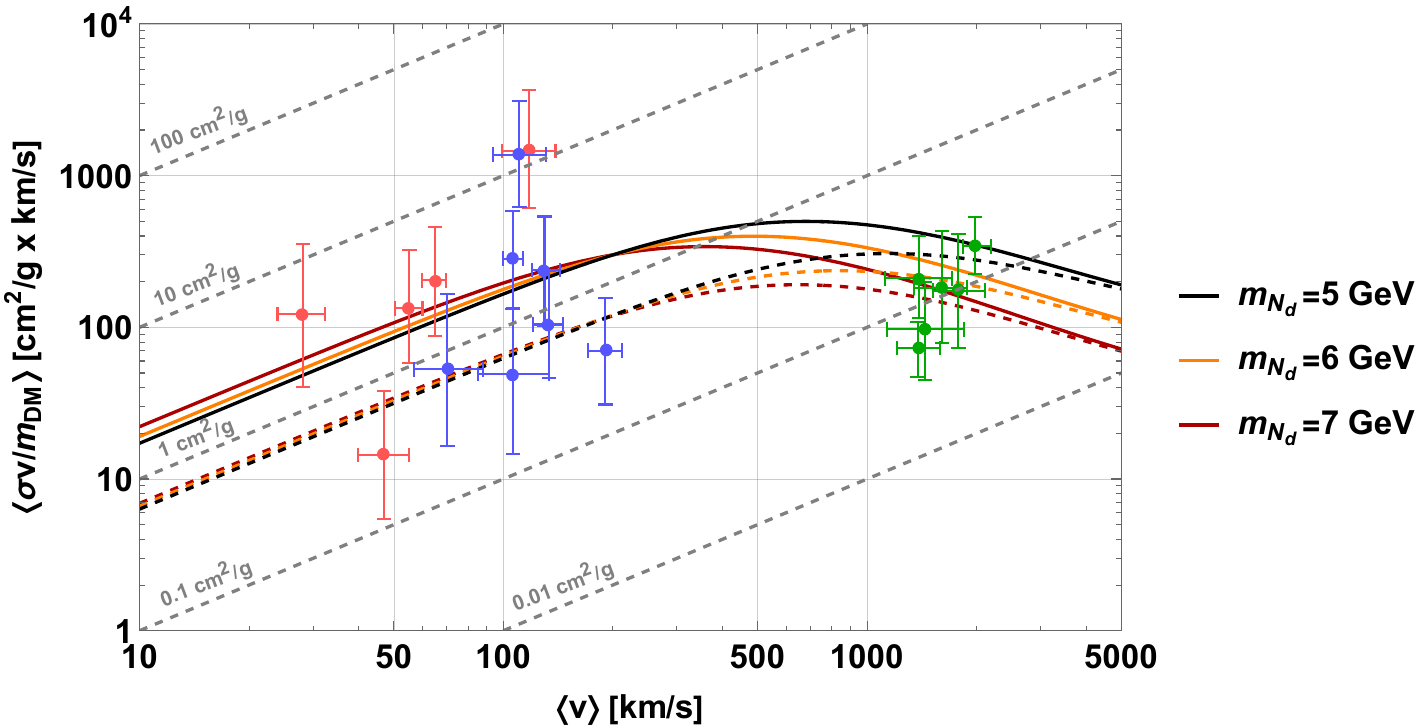}
	\caption{The velocity-weighted self-interaction cross sections per unit DM mass as a function of mean DM collision velocity. The points, obtained from Ref.~\cite{Kaplinghat:2015aga}, correspond to discrepancies between simulations and observations from various astrophysical sources: dwarf galaxies (red), LSBs (blue) and galaxy clusters (green). The diagonal lines represent contours of constant $ \sigma/m_{\rm DM} $ (cross section per unit mass). For different DM masses $m_{N_d} = 5, 6, 7$~GeV, the solid (dashed) lines depict the predictions from the dark baryon model, considering $ \sigma/m_{\rm DM}=1.5(0.6)~\text{cm}^2/\text{g} $ for $ \langle v\rangle =200 $~km/s. In these cases, the dark pion mass is $ m_{\pi_d}\sim 0.15 m_{N_d} $ (i.e. $ m_{\pi_d}\sim 0.58 \Lambda_{\rm dQCD} $). }
	\label{fig: SIDM}
\end{figure}

In Figure~\ref{fig: SIDM}, velocity-weighted self-interaction cross sections (given by Eq.~(\ref{eq: velocity-weighted CS})) per unit DM mass are depicted for varying mean DM collision velocity. The points, sourced from Ref.~\cite{Kaplinghat:2015aga}, hint at discrepancies between simulations and observations derived from diverse astrophysical sources: dwarf galaxies (shown as red points), low surface brightness (LSB) spiral galaxies (blue points) and galaxy clusters (green points). The diagonal lines within the plot represent contours of constant $ \sigma/m_{\rm DM} $, indicating the cross section per unit DM mass. The solid (dashed) lines show the predictions from this dark baryon model for different DM masses ($m_N = 5, 6, 7$ GeV), where $ \sigma/m_{\rm DM}=1.5(0.6)\text{cm}^2/\text{g} $ at $ \langle v\rangle =200 $~km/s, corresponding to an example involving the Milky Way. 
The figure illustrates that it is possible to attain DM self-interaction cross sections that exhibit the necessary velocity-dependence to account for astrophysical observations ranging from dwarf galaxies to galaxy clusters when the DM particle has a mass of approximately 5 GeV.

In all the depicted examples within the figure, the dark pion mass is approximately $m_{\pi_d}\sim 0.15 m_{N_d}$ (or $m_{\pi_d}\sim 0.58 \Lambda_{\rm dQCD}$). By utilizing the estimation $m_{\pi_d}\propto\sqrt{m_{q_d}\Lambda_{\rm dQCD}}$, we find that $ m_{q_d}/m_q \sim  6, 7,9 $ for $m_{N_d} = 5, 6, 7$~GeV (or $\Lambda_{\rm dQCD}=1.3, 1.6, 1.9$~GeV), respectively. Since the light quark masses in QCD are around $ 2-5 $~MeV, the dark quark masses are $ m_{q_d} \sim 12-45 $~MeV for the above examples. As a result, we observe that $m_{q_d}\ll \Lambda_{\rm dQCD}$, which suggests the possibility of a first-order phase transition induced by dark confinement. To obtain a first-order phase transition, the dark strange quark is also required to be light as minimum according to Ref.~\cite{Pisarski:1983ms}. In the paper~\cite{Cuteri:2021ikv}, they conducted lattice investigations to examine whether a first-order phase transition happens or not when taking the chiral limit ($ m_{q_d}\rightarrow 0 $) and having a minimum of three quarks. These studies indicate that the region where such a phase transition takes place might be narrower than initially expected. However, it is crucial to acknowledge that these investigations highlight the inherent uncertainties associated with lattice calculations in this particular limit.

\section{New $1-10 $ TeV physics}
\label{sec: new physics}

In this section, we will explore the potential connection between the concept of dark baryons at the GeV scale and the existence of new physics within the range of $ \mathcal{O}(1-10) $ TeV. Firstly, we will outline three theoretical frameworks that have been developed to explain the comparable masses of VM and DM particles. The implications of these theories suggest the existence of new particles with masses on the order of $1-10$ TeV. Secondly, we emphasize the significance of the $\mathcal{O}(1-10)$ TeV-scales as a relevant range for addressing the EW hierarchy problem, through the introduction of either composite Higgs or SUSY. These theoretical frameworks can give rise to intriguing collider signatures and the possibility of detecting GW signals at frequencies around the mHz range, originating from phase transitions associated with the Higgs condensation or SUSY breaking. Such GW signals are testable with future GW detectors like LISA. Finally, there is compelling motivation to suggest that both VM and DM asymmetries are generated through leptogenesis at a scale of approximately $ 1-10 $ TeV.

\subsection{The mass-density coincidence}
\label{sec: mass-density coincidence}

From now on, we interpret the cosmological mass density coincidence between DM and VM, $ \Omega_{\rm DM}\simeq 5 \Omega_{\rm VM} $, as a hint towards a fundamental connection between the origins of DM and VM abundances. However, majority of prominent DM candidates do not provide a clear explanation for this relationship. The number density of VM is attributed to the baryon asymmetry of the universe, while its mass arises from the confinement energy of QCD. Given these distinct mechanisms for generation, there is no inherent reason why the cosmological mass densities of these species should align within the same order of magnitude. This coincidence in the mass densities largely motivates a DM particle-antiparticle asymmetry, similar to what is observed for VM. Referred to as the ADM hypothesis, these related asymmetries likely arose from specific processes during an early cosmological epoch that subsequently decoupled, resulting in similar number densities for VM and DM --- $ n_{\rm VM} \sim n_{\rm DM} $. 

In the following, we outline three theoretical frameworks that address the connection between DM and VM masses and elucidate the apparent coincidence between the cosmological mass densities of DM and VM, $ \Omega_{\rm DM}\simeq 5 \Omega_{\rm VM} $. Inspired by the analogy with the proton in QCD, we explore ADM candidates that emerge as confined states within a dark QCD-like gauge group. This endeavor entails devising a mechanism to establish a relationship between the confinement scales of the dark gauge group and visible QCD, resulting in $ \Lambda_{\rm dQCD} /\Lambda_{\rm QCD}\sim 5 $. 
Previous attempts to establish this correlation have pursued the following avenues of inquiry: \begin{itemize}
	\item[(i)] Introduction of a symmetry between the gauge groups often gives rise to intriguing models of mirror matter, wherein the mirror symmetry is either preserved~\cite{Hodges:1993yb,Foot:2003jt,Foot:2004pq,Foot:2004pa,Ibe:2018juk} or deliberately broken in a well-considered manner~\cite{An:2009vq,Cui:2011wk,Akamatsu:2014qsa,Farina:2016ndq,Lonsdale:2017mzg,Lonsdale:2018xwd,Ibe:2019ena,Beauchesne:2020mih,Feng:2020urb,Ritter:2021hgu,Ibe:2021gil,Bodas:2024idn}. Twin Higgs models serve as a notable illustration of such scenarios~\cite{Farina:2015uea,Craig:2015xla,GarciaGarcia:2015pnn,Barbieri:2016zxn}. They attract attention due to their ability to ameliorate the little hierarchy problem by attempting to explain the gap between the EW scale and the scale of either CH or SUSY models. 
	\item[(ii)] Extending the field content enables a modification that leads to the convergence of the running gauge couplings in visible and dark QCD towards infrared (IR) fixed points with similar magnitudes, as originally proposed in Ref.~\cite{Bai:2013xga} and further explored in Refs.~\cite{Newstead:2014jva,Ritter:2022opo}. As a consequence, the confinement scales of visible and dark QCD become related through this dynamical mechanism, resulting in $ m_{\rm VM} \sim m_{\rm DM} $. 
	\item[(iii)] The concept of dark unification, as proposed in Ref.~\cite{Murgui:2021eqf}, involves the realization of a unified framework encompassing both dark and visible sectors. In this framework, the masses of dark to visible baryons are determined by the ratio of the dark to visible confinement scales. These confinement scales, in turn, are found to be closely matched in magnitude due to the unification of the dark and visible gauge theories at a high energy scale. 
\end{itemize} 

For these three frameworks, the mechanisms generally require particles at the $ \mathcal{O}(1-10) $ TeV energy scale, as outlined below: \begin{itemize}
	\item[(i)] If a mirror symmetry is introduced, a notable indication would be the observation of the mirror Higgs boson itself, with a mass around $ \mathcal{O}(1-10) $ TeV. Its decay into $ WW $ and $ ZZ $ particles could serve as a compelling signature that can potentially be detected at the High-Luminosity LHC (HL-LHC) or future colliders.
	\item[(ii)] In the context of the IR fixed-point approach, it is noted that these models usually require the mass scale of the new particle content to be in the vicinity of the TeV range for a wide range of initial couplings at the ultraviolet (UV) scales. As a result, these models are subject to stringent constraints from collider experiments.	
	\item[(iii)] In the case of dark unification, collider experiments offer the potential to observe distinctive signals arising from so-called connector states. These states act as mediators for darko-baryo-genesis and establish a vital connection between the dark and visible baryon sectors. Their masses typically fall within the range of $\mathcal{O}(1-10)$ TeV. At colliders, these connector states can be produced in pairs and subsequently decay into SM jets accompanied by dark sector states.	\end{itemize} While the detection of these particles at the LHC remains elusive, there exists the possibility of their discovery at HL-LHC or in future particle colliders. This provides motivation for the construction of more powerful particle colliders capable of exploring higher energy regimes. 

\subsection{Generation of asymmetries}
\label{sec: Generation of asymmetries}

To explain the relation between visible and dark baryon asymmetries, i.e. $ n_{\rm VM}\sim n_{\rm DM} $, various models are construced based on the concept of leptogenesis~\cite{Fukugita:1986hr}. Leptogenesis is a well-known mechanism that explains the baryon asymmetry in the Universe (BAU) --- typically quantified by the cosmic baryon-to-photon ratio $ \eta_B^{\rm obs}\simeq 6.1\times 10^{-10} $~\cite{ParticleDataGroup:2016lqr,Planck:2015fie}. It is closely linked to the type-I seesaw mechanism~\cite{Minkowski:1977sc,Sawada:1979dis,Yanagida:1980xy,Gell-Mann:1979vob,Mohapatra:1979ia}, which introduces sterile right-handed neutrinos ($N_i$) with large Majorana masses ($M_i$) to explain the small masses of the SM neutrinos. In the standard thermal leptogenesis, the heavy right-handed neutrinos are produced through scatterings in the thermal bath. Then, their CP-violating out-of-equilibrium decays generate a primordial lepton asymmetry. This lepton asymmetry is later converted into a baryon asymmetry through EW sphaleron processes. Apart from the lepton asymmetry, it is also possible to generate asymmetries in other quantum numbers from the decays of $N_i$~\cite{An:2009vq,Falkowski:2011xh}, leading to a natural equivalence between number densities $n_{\rm VM}$ and $n_{\rm DM}$. 

However, standard thermal leptogenesis has a limitation in that it requires a very high mass scale for the right-handed neutrinos, where the lower bound on the lightest one is $ M_i^{\rm min}\gtrsim 10^8 $~GeV~\cite{Blanchet:2008pw}. This high mass scale poses several challenges. Firstly, it prevents the direct exploration of leptogenesis dynamics in future collider experiments~\cite{Chun:2017spz}. Secondly, the detection of lepton number violation at low energies could potentially rule out high-scale leptogenesis~\cite{Deppisch:2013jxa,Deppisch:2015yqa,Harz:2015fwa}. Thirdly, in the type-I seesaw model, the presence of heavy right-handed neutrinos affects the renormalization group running of the SM Higgs mass parameter, requiring fine-tuning at the EW scale, which leads to a naturalness problem~\cite{Vissani:1997ys}. Finally, due to the fact that we have motivated a new physics scale at approximately $\mathcal{O}(1-10)$ TeV, low-scale alternatives to standard thermal leptogenesis are very motivated, such as resonant leptogenesis~\cite{Pilaftsis:1997jf,Pilaftsis:2003gt,Suematsu:2011va} and thermal leptogenesis in E. Ma's scotogenic model of radiative neutrino masses~\cite{Ma:2006km,Cai:2017jrq} or other radiative neutrino mass models~\cite{Cai:2017jrq}. Interestingly, radiative neutrino mass generation can be realized with
composite Higgses~\cite{Cacciapaglia:2020psm,Rosenlyst:2021tdr}. These alternatives allow for masses of the new states to be around $\mathcal{O}(1-10)$ TeV. Finally, another possibility to address the observed baryon asymmetry of the universe, as demonstrated in Refs.~\cite{Espinosa:2011eu,Bruggisser:2018mus,Bruggisser:2018mrt,Xie:2020bkl,Bruggisser:2022rdm,vonHarling:2023dfl}, involves EW baryogenesis in the context of CH models.

\subsection{UV-complete realizations}
\label{sec: UV-complete realizations}

The EW hierarchy problem has been a focus of intense research, predating the discovery of the Higgs boson at the LHC in 2012. The instability of the Higgs mass to large radiative corrections continues to motivate the search for new physics around the TeV scale. Various theories propose explanations for the observed Higgs mass without excessive fine-tuning. These include UV-complete models like Natural SUSY~\cite{Batra:2003nj,Ellwanger:2009dp,Hall:2011aa,Curtin:2014zua,Casas:2014eca} and CH~\cite{Kaplan:1983sm}, as well as proposals like the Little Higgs~\cite{Arkani-Hamed:2001nha,Schmaltz:2005ky} that stabilize the Higgs sector up to an experimental cutoff of $ 5-10 $ TeV, with additional physics ensuring naturalness at higher scales. Interestingly, in Ref.~\cite{Frandsen:2011kt}, it is demonstrated that a $\sim 5$ GeV dark baryon, similar to the one considered here, with a cosmic asymmetry akin to that of baryons, serves as a natural DM candidate and may arise from the same strong gauge group of CH dynamics.

However, these solutions predict the existence of TeV-scale particles with quantum numbers similar to the SM, particularly QCD color charges. Despite expectations, these particles have not yet been detected at the LHC, prompting the need for more powerful particle colliders. This discrepancy gives rise to the ``little hierarchy problem.'' To address this tension, a class of models called neutral naturalness models~\cite{Chacko:2005pe,Chacko:2005un,Burdman:2006tz,Craig:2014aea,Craig:2015pha,Barbieri:2016zxn,Serra:2017poj,Xu:2018ofw} has been developed. In these models, partner particles are not charged under SM color, achieved by a symmetry that does not commute with SM color. These models, such as the Twin Higgs models~\cite{Chacko:2005pe,Craig:2015pha,Craig:2014aea,Barbieri:2016zxn,Serra:2017poj,Xu:2018ofw}, solve the little hierarchy problem by canceling divergent loop corrections to the Higgs potential through the introduction of a twin sector related to the SM through a discrete $ \mathbb{Z}_2 $ symmetry.

These model frameworks also appear to be potentially testable through GWs at the mHz frequency range. These GWs may be generated by potential strong first-order EW phase transitions in CH models, as studied in Refs.~\cite{Bian:2019kmg,Xie:2020bkl,Frandsen:2023vhu}, or from broken SUSY, as discussed in Refs.~\cite{Apreda:2001tj,Craig:2020jfv}. Future GW detectors such as LISA may provide an avenue for testing these scenarios. In Section~\ref{sec: Gravitational waves generated in dark baryon models}, we will explore the prospects of testing these dark baryon models by GWs with frequencies in vicinity of $\mu\text{Hz}$. As a result, the models discussed in this paper have the capability of generating GWs with both frequencies approximately around $\mathcal{O}(\mu\text{Hz})$ and $\mathcal{O}(\text{mHz})$. These GWs originate from first-order phase transitions potentially generated by the dark baryon sector and a UV-complete model (such as CH or SUSY models) at approximately $\mathcal{O}(1-10)$ TeV.

\section{Gravitational waves generated in dark baryon models}
\label{sec: Gravitational waves generated in dark baryon models}

In this section, we unveil the interesting possibility that a first-order phase transition induced by dark confinement at around $\mathcal{O}(1)$ GeV could potentially generate GWs that almost span the frequency gap, extending from approximately $\mu$Hz to mHz. This finding serves as a strong motivation for the advancement of GW experiments capable of bridging this frequency gap in the landscape of GW research. In the following, we introduce the necessary formalism for computing the signals of these GWs.

\subsection{Formalism of gravitational-wave signals}
\label{sec: Gravitational waves}

During a first-order phase transition, GWs can be generated by bubbles collisions~\cite{Huber:2008hg,Jinno:2016vai}, stirred acoustic waves~\cite{Hindmarsh:2015qta,Hindmarsh:2017gnf} and magnetohydrodynamic turbulence in the super-cooling plasma~\cite{Caprini:2009yp,Binetruy:2012ze}. The GW spectrum is given by \begin{equation} \begin{aligned} \label{eq: total GW power spectrum}
\Omega_{\rm GW}(\nu)=\frac{1}{\rho_c}\frac{d\rho_{\rm GW}}{d\ln \nu}\,,
\end{aligned}\end{equation} where $ \rho_c = 3H_0^2 /(8\pi G) $ is the critical energy density today, $ \nu $ denotes the frequency of the GWs and $\rho_{\rm GW} $ is the GW energy density produced during first-order phase transitions. Here, $H_0 = 100 h~\text{km~s}^{-1}\text{Mpc}^{-1} $ is the Hubble rate today with $h=0.72$. 

The GW spectrum from first-order phase transitions is generally characterized by three essential parameters, $ \alpha $, $ \beta/H_n $ and the wall velocity of the first-order phase transition bubble $ v_w $, evaluated at the nucleation temperature, $ T_n $. The parameter $ \alpha $ is the ratio of the vacuum energy density $ \epsilon $ and radiation energy density $ \rho_{\rm rad} $, while $ \beta /H_n $ is the nucleation rate divided with the Hubble rate evaluated at $ T_n $ which measures the time duration of the phase transition. Thus, the smaller $ \beta /H_n $ is, the stronger the phase transition is. These parameters are, therefore, given by \begin{equation} \begin{aligned}\label{eq: alpha beta/Hn}
\alpha \equiv \frac{\epsilon}{\rho_{\rm rad}}\,,   \quad\quad \frac{\beta}{H_n}\equiv T_n \frac{d}{dT}\left(\frac{S_3}{T}\right)\bigg \vert_{T=T_n}\,,
\end{aligned}\end{equation} where \begin{equation} \begin{aligned}
\epsilon=&-\Delta V_T +T_n\Delta\frac{\partial V_T}{\partial T} \bigg\vert_{T=T_n}\,, \quad\quad \rho_{\rm rad}= \frac{\pi^2}{30}g_* T_n^4\,, \quad\quad \beta=\frac{d}{dt}\left(\frac{S_3}{T}\right)\bigg \vert_{t=t_n}\,.
\end{aligned}\end{equation} Here, ``$ \Delta $'' denotes the difference between the true and false vacua, while $ g_* $ and $ t_n $ are the relativistic degrees of freedom and the cosmic time at $ T_n $, respectively. 

The total power spectrum of the GWs in Eq.~(\ref{eq: total GW power spectrum}) consists of the three above-mentioned components, which can be written as \begin{equation} \begin{aligned}\label{eq: total power spectrum}
\Omega_{\rm GW}(\nu)h^2=\Omega_{\rm env}(\nu)h^2+\Omega_{\rm sw}(\nu)h^2+\Omega_{\rm turb}(\nu)h^2\,, 
\end{aligned}\end{equation} where $ \Omega_{\rm env}(\nu)h^2 $, $ \Omega_{\rm sw}(\nu)h^2 $ and $ \Omega_{\rm turb}(\nu)h^2 $ are the power spectra of the GWs, respectively, from bubble collisions, sound waves and turbulence of the magnetohydrodynamics in the particle bath. The full expressions of these spectra are given in Refs.~\cite{Ellis:2018mja, Ellis:2020awk}. These expressions are modified compared to them given by e.g. Ref.~\cite{Chen:2017cyc}, since the sound wave period does not last a full Hubble time. This results in a reduction of the total GW signal. In Refs.~\cite{Ellis:2019oqb, Ellis:2020nnr}, it is further shown that bubble collisions are
almost negligible in transitions with polynomial potentials. These issues are also reviewed by LISA in Ref.~\cite{Caprini:2019egz}. Therefore, we will in the following neglect these contributions, i.e. $\Omega_{\rm GW}(\nu)\approx \Omega_{\rm sw}(\nu)+\Omega_{\rm turb}(\nu) $. According to Ref.~\cite{Ellis:2020awk}, the sound wave contribution can be written as \begin{equation} \begin{aligned}\label{eq: sw GW power spectrum}
\Omega_{\rm sw}(\nu)h^2 \simeq& 4.18\times 10^{-6} v_w (H_n\tau_{\rm sw}) \left(\frac{100}{g_*}\right)^{1/3} \left(\frac{\beta}{H_n}\right)^{-1} \left(\frac{\kappa_{\rm v}\alpha}{1+\alpha}\right)^2 \left(\frac{\nu}{\nu_{\rm sw}}\right)^3 \\&\times \left[1+\frac{3}{4}\left(\frac{\nu}{\nu_{\rm sw}}\right)^2\right]^{-7/2}\,,
\end{aligned}\end{equation} where the peak frequency is given by \begin{equation} \begin{aligned}\label{eq: sw peak frequency}
\nu_{\rm sw}=1.91\times 10^{-5}~\text{Hz}\left(\frac{T_n}{100~\text{GeV}}\right)(1+\alpha)^{1/4}\left(\frac{g_*}{100}\right)^{1/6}\frac{1}{v_w}\left(\frac{\beta}{H_n}\right)\,,
\end{aligned}\end{equation} while the duration of the sound wave source can be approximated as follows~\cite{Ellis:2020awk} \begin{equation} \begin{aligned}\label{eq: sound wave period }
H_n \tau_{\rm sw}\approx \text{min}\left[1, (8\pi)^{1/3}\text{max}(v_w,c_s) \sqrt{\frac{4}{3}\frac{1+\alpha}{\kappa_{\rm v}\alpha}}\left(\frac{\beta}{H_n}\right)^{-1}\right]\,.
\end{aligned}\end{equation} Here $c_s$ is the speed of sound in the plasma, given by $c_s=\sqrt{1/3}$ in a relativistic fluid, and $\kappa_{\rm v}$ is the ratio of the bulk kinetic energy to the vacuum energy, fitted in Ref.~\cite{Espinosa:2010hh} as function of $\alpha$ and $v_w$. The turbulence contribution can be written as~\cite{Ellis:2020awk} \begin{equation} \begin{aligned}\label{eq: turb GW power spectrum}
\Omega_{\rm turb}(\nu)h^2 \simeq & 3.32\times 10^{-4} v_w (1-H_n\tau_{\rm sw}) \left(\frac{100}{g_*}\right)^{1/3} \left(\frac{\beta}{H_n}\right)^{-1} \left(\frac{\kappa_{\rm v}\alpha}{1+\alpha}\right)^{2/3} \left(\frac{\nu}{\nu_{\rm turb}}\right)^3 \\&\times \left[1+\frac{\nu}{\nu_{\rm turb}}\right]^{-11/3}\left[1+\frac{43.98}{v_w}\left(\frac{\beta}{H_n}\right) \left(\frac{\nu}{\nu_{\rm turb}}\right)\right]^{-1}\,,
\end{aligned}\end{equation} where the peak frequency is given by \begin{equation} \begin{aligned}\label{eq: turb peak frequency}
\nu_{\rm turb}=2.89\times 10^{-5}~\text{Hz}\left(\frac{T_n}{100~\text{GeV}}\right)(1+\alpha)^{1/4}\left(\frac{g_*}{100}\right)^{1/6}\frac{1}{v_w}\left(\frac{\beta}{H_n}\right)\,.
\end{aligned}\end{equation} In the following calculations, the significant contribution source
of the GWs is from the sound waves, while the turbulence contribution is a subleading source for the GWs. 

\subsection{Bridging the $ \mu $Hz gap in the gravitational-wave landscape}
\label{sec: Bridging the gap in the gravitational-wave landscape}

In this section, we will discuss the primary objective of this paper, focusing on the calculations of the potential GW signals ($\Omega_{\rm GW}(\nu)\approx \Omega_{\rm sw}(\nu)+\Omega_{\rm turb}(\nu) $) originating from dark baryon models, as discussed in Section~\ref{sec: Dark baryons}. Specifically, we use the expressions in Eqs.~(\ref{eq: sw GW power spectrum}) and~(\ref{eq: turb GW power spectrum}) to compute the GW contributions arising from sound waves and turbulence, respectively. We emphasize that if the phase transition triggered by dark confinement, occurring around $\mathcal{O}(1)$~GeV, is of first order, it has the potential to generate GWs covering a wide frequency range, from $\mu$Hz to mHz. Notably, the field of GW research has largely left the $\mu$Hz frequency band unexplored.

In Ref.~\cite{He:2022amv}, they have studied GWs and primordial black hole (PBH) productions from a holographic model describing the gluon sector of Yang-Mills theories. They state that they are not able to compute the parameter $ \beta/H_n $ from first principles, but it can be constrained by the PBH abundance associated with the first-order phase transition with $ \beta/H_n >8.59$. This constraint results in a GW peak frequency around $ \mu\text{Hz} $. Commonly, this parameter is estimated to be approximately $ \beta/H_n \approx 4\log(m_{\rm Pl}/T_n) \approx \mathcal{O}(100) $~\cite{Hogan:1983ixn,Hogan:1986qda}, where $ m_{\rm Pl} $ denotes the Planck mass and the nucleation temperature is approximately $ T_n \approx \mathcal{O}(1) $ GeV in our cases. It is worth noting that the second relation remains valid for a broad range of $ T_n $ values. However, there are studies (with dark quarks~\cite{Helmboldt:2019pan} and purely gluonic~\cite{Reichert:2022naa,Morgante:2022zvc}) suggesting that $ \beta/H_n \sim \mathcal{O}(10^4)$ based on various effective models used to calculate these GW parameters. Consequently, this yields GW peak frequencies in the mHz ballpark, which unfortunately fall below the sensitivity range of future experiments like LISA. It is important to acknowledge that these calculations have inherent uncertainties. To obtain precise quantitative results, true first-principle calculations, such as lattice simulations, are necessary. However, it is worth noting that these simulations have primarily been attempted in the context of pure $ \SU(N) $ Yang-Mills theory~\cite{Huang:2020crf}, without the inclusion of dark quarks. However, regardless of the specific value of $ \beta/H_n $ falling within the range $ 8.59 < \beta/H_n \lesssim \mathcal{O}(10^4)$, the resulting peak frequency typically lies within the frequency gap that is poorly covered by common future experiments. Therefore, in the subsequent calculations, we will scan over the parameter range of $ \beta/H_n $ from $ 10 $ to $ 10^5 $ to encompass all the relevant studies.

\begin{figure}[t]
	\centering
	\includegraphics[width=0.8\textwidth]{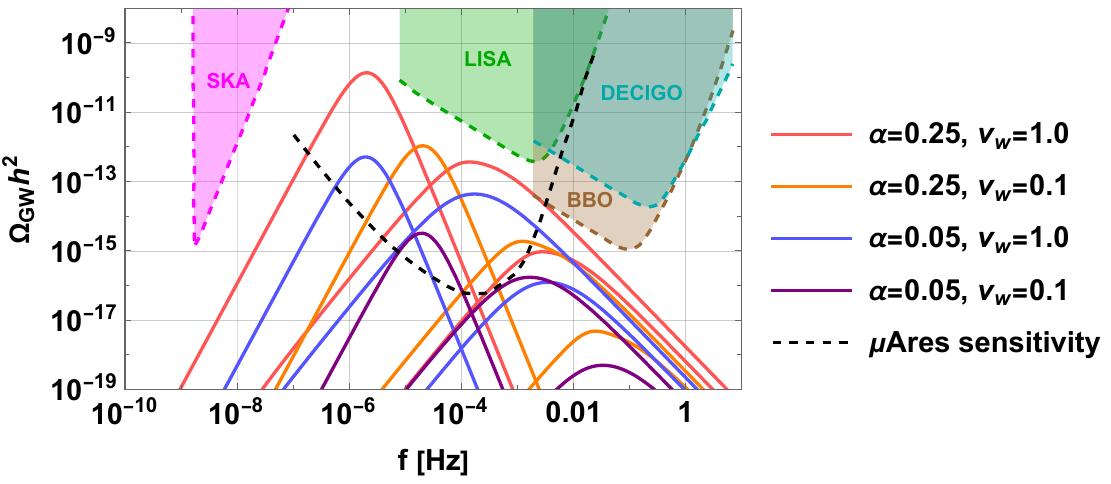}
	\caption{The GW signals as a function of frequency for various parameters in the scenario involving a $ \sim 5 $ GeV dark baryon, where the nucleation temperature is set at $ T_n = 1 $ GeV. The colored lines are based on different values of $ \alpha $ and $ v_w $. The lines with the same color, from left to right, represent $ \beta/H_n =10 $, $ 500 $ and $ 10^4 $, respectively. The shaded regions represent the sensitivities for various future GW detectors such as LISA, BBO, DECIGO and SKA, while the black dashed line sets the sensitivity of the proposed GW detector $ \mu $Ares~\cite{Sesana:2019vho}. }
	\label{fig: figure 2}
\end{figure}

Figure~\ref{fig: figure 2} depicts the GW signals plotted as function of frequency for different parameter values, where the nucleation temperature is assumed to be $ T_n = 1 $ GeV. The colored lines correspond to different combinations of $ \alpha $ and $ v_w $. Within each color group, the lines from left to right represent $ \beta/H_n =10 $, $ 500 $ and $ 10^4 $, respectively. The shaded regions indicate the sensitivities of future GW detectors, including LISA, BBO, DECIGO, and SKA. According to these calculations, the generated GWs span a wide frequency range, from $\mu$Hz to mHz, which is poorly covered by common future detectors. This finding provides compelling grounds for pursuing the development of GW experiments capable of closing this frequency gap in the GW landscape. Consequently, this serves as a strong motivation for advancing research and technology in the field of GWs. Furthermore, it is worth noting that interpreting dark mesons or glueballs from dark confinement as DM~\cite{Kribs:2016cew,Cline:2021itd} will lead to similar conclusions.

\begin{figure}[t]
	\centering
\includegraphics[width=0.75\textwidth]{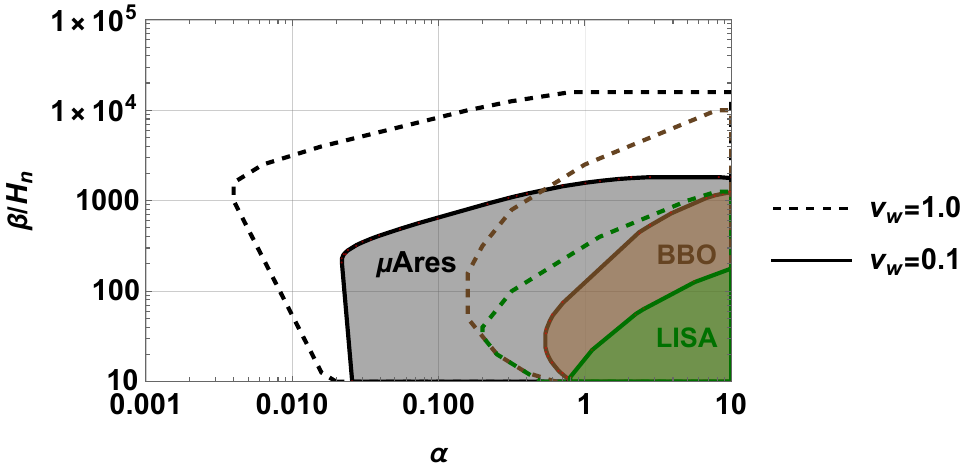}
	\caption{The constraints on the $ \alpha - \beta/H_n $ parameter space in the scenario where the nucleation temperature is $ T_n = 1 $ GeV, determined by the sensitivities of future GW detectors such as LISA, BBO and $ \mu $Ares~\cite{Sesana:2019vho}. Dashed lines represent $ v_w = 1.0 $, while solid lines represent $ v_w = 0.1 $. The enclosed regions correspond to the parameter space potentially covered by these experiments. }
	\label{fig: figure 3}
\end{figure}

An example of a future experiment that could fill this frequency gap is the proposed space-based interferometer called $ \mu $Ares~\cite{Sesana:2019vho}, surveying GWs within the frequency range of $ \mu $Hz to mHz. In Figure~\ref{fig: figure 2}, the black dashed line corresponds to the sensitivity of this proposed GW detector. This detector will possess high sensitivity, allowing exploration of a significantly larger portion of the parameter space associated with GWs originating from these dark baryon models. Additionally, Figure~\ref{fig: figure 3} illustrates constraints imposed on the parameter space of $ \alpha $ and $ \beta/H_n $, derived from the sensitivities of future detectors, including LISA, BBO and $ \mu $Ares.~\footnote{It is important to note that constraints from DECIGO and SKA have not been included into Figure~\ref{fig: figure 3}, as they either weakly constrain the parameter space or do not impose any constraints.} The dashed lines represent the case where $v_w = 1.0$, while solid lines correspond to $v_w = 0.1$. The enclosed regions within the plot depict the parameter space potentially accessible to these experimental setups. The figure shows that the $ \mu $Ares detector exhibits the potential to cover a substantial part of the parameter space, rendering it an intriguing experiment to pursue.

To summarize, this study presents compelling reasons to advance the development of GW experiments that can bridge the frequency gap in the GW landscape, specifically focusing on the range from $ \mu $Hz to mHz like the $\mu$Ares experiment. This discovery serves as a powerful motivation to propel progress in GW research and technology.

\section{Conclusions}
\label{sec: Conclusions}

In this paper, we have studied gravitational waves (GWs) with frequencies in the $\mu$Hz ballpark, originating from dark confinement in dark baryon models. We assume the existence of a dark version of Quantum Chromodynamics, predicting a $ \sim 5~\text{GeV} $ dark baryon (analogous to ordinary baryons), which serves as Dark Matter (DM). Given that the present-day mass densities of DM and Visible Matter (VM) are of the same order, we consider the Asymmetric DM hypothesis. This hypothesis suggests that the abundance of DM shares the same origin as VM: a particle-antiparticle asymmetry in number densities. This similarity in mass densities also implies that the DM and VM particles may have comparable masses of $ \mathcal{O}(\text{GeV}) $. Thus, we outline three approaches that account for this, including mirror symmetry~\cite{Hodges:1993yb,Foot:2003jt,Foot:2004pq,Foot:2004pa,Ibe:2018juk,An:2009vq,Cui:2011wk,Akamatsu:2014qsa,Farina:2016ndq,Lonsdale:2017mzg,Lonsdale:2018xwd,Ibe:2019ena,Beauchesne:2020mih,Feng:2020urb,Ritter:2021hgu,Ibe:2021gil,Farina:2015uea,Craig:2015xla,GarciaGarcia:2015pnn,Barbieri:2016zxn}, convergence of running gauge couplings towards infrared fixed points~\cite{Bai:2013xga} and the concept of dark unification~\cite{Murgui:2021eqf}. We emphasize that these mechanisms, in general, predict the existence of new TeV-mass particles.

Therefore, we explored the potential connection between the concept of dark baryons at the GeV scale and the existence of new physics within the range of $ \mathcal{O}(1-10) $ TeV. In addition to predicting new TeV-mass particles from the three theoretical frameworks addressing the comparable masses of VM and DM, we emphasize the significance of the $\mathcal{O}(1-10)$~TeV scales. These scales are pertinent for addressing the electroweak hierarchy problem, either through the introduction of Composite Higgs~\cite{Kaplan:1983sm} or Supersymmetry (SUSY)~\cite{Batra:2003nj,Ellwanger:2009dp,Hall:2011aa,Curtin:2014zua,Casas:2014eca}. These theoretical frameworks can give rise to intriguing TeV collider signatures and the possibility of detecting GW signals at frequencies around the mHz range, arising from phase transitions associated with the Higgs condensation~\cite{Bian:2019kmg,Xie:2020bkl,Frandsen:2023vhu} or SUSY breaking~\cite{Apreda:2001tj,Craig:2020jfv}. Such GW signals are testable with future GW detectors like LISA. Finally, there is compelling motivation to suggest that both VM and DM asymmetries are generated through leptogenesis at a scale of approximately $ 1-10 $ TeV, such as resonant leptogenesis~\cite{Pilaftsis:1997jf,Pilaftsis:2003gt,Suematsu:2011va} and thermal leptogenesis in E. Ma's scotogenic model of radiative neutrino masses~\cite{Ma:2006km,Cai:2017jrq} or other radiative neutrino mass models~\cite{Cai:2017jrq}. Interestingly, radiative neutrino mass generation can be realized with composite Higgses~\cite{Cacciapaglia:2020psm,Rosenlyst:2021tdr}. These models provide masses of the new states to be around $\mathcal{O}(1-10)$ TeV, making them testable in future colliders.

In conclusion, this study serves as a strong motivation for advancing GW experiments capable of bridging the $\mu$Hz frequency gap in the GW landscape. Simultaneously, the construction of more powerful particle colliders for probing higher energy regimes, especially physics at $\mathcal{O}(1-10)$ TeV scale, is highly motivated by this work. Given the feasibility of testing these models from various perspectives, we strongly advocate for the further development of these model frameworks.


\section*{Acknowledgements}
I would like to thank Prateek Agrawal for insightful discussions. I acknowledge funding from The Independent Research Fund Denmark, grant number DFF 1056-00027B.  

%
\bibliography{GW_DB.bib}
\bibliographystyle{JHEP}
%

\end{document}